\documentclass[12pt]{article}
\usepackage{graphicx,comment,amsfonts,amsmath}

\def\E{{\sf E}}

\def\ur{\underline{r}}
\def\Rcal{\mathcal{R}}

\def\eg{{\em e.g.},~}
\def\ie{{\em i.e.},~}
\def\cf{{\em cf.},~}
\newtheorem{proposition}{Proposition}[section]

\newtheorem{theorem}{Theorem}[section]
\newtheorem{corollary}{Corollary}[section]
\newcommand{\beqa}{\begin{eqnarray*}}
\newcommand{\eeqa}{\end{eqnarray*}}
\newcommand{\be}{\begin{eqnarray}}
\newcommand{\ee}{\end{eqnarray}}

\renewcommand{\th}[1]{{#1}^{\rm th}}

\newcommand{\Minv}[1]{M^{-1}_{#1}}

\newcommand{\Hinv}[1]{H^{-1}_{#1}}

\begin{document}



\title{Stationary Distribution of a\\
 Generalized LRU-MRU Content Cache\thanks{This 
research supported in part by a Cisco Systems URP gift and 
NSF CNS grant 1526133.}}

\author{George Kesidis\\
School of EECS\\
Pennsylvania State University\\
University Park, PA, 16802, USA\\
Email: gik2@psu.edu
}

\maketitle

\begin{abstract}
Many different caching mechanisms have been previously proposed,
exploring different insertion and eviction policies and their performance
individually and as part of caching networks.
We obtain a 
novel closed-form stationary invariant distribution 
for a generalization  of
Least Recently Used (LRU) 
and 
Most Recently Used
(MRU) eviction for single caching nodes under
a reference Markov model.
Numerical comparisons are made with an 
``Incremental Rank Progress" (IRP a.k.a. CLIMB) 
and random eviction (RE a.k.a. random replacement, RANDOM) methods
under a steady-state
Zipf popularity distribution.
The range of cache hit probabilities is smaller under MRU and
larger under IRP compared to LRU. We conclude with
the invariant distribution for a special case of a 
RE caching tree-network.
\end{abstract}

\section{Introduction}

Caching is a ubiquitous mechanism in communication and computer
systems. 
The role of a content caching network is to reduce the load on the
origin servers of requested data objects, 
reduce the required network bandwidth to transmit content\footnote{That is,
content that is not encrypted for particular end-users.},
and reduce the response times to the queries. 
Caching in computational settings reduces
 delays associated with disk IO (page caches).
Data actively being, or likely
soon to be, accessed by  a CPU is stored in 
lower-level caches, \ie  memories closer (with less access time)
to the CPU.

The invariant distribution of 
the widely deployed  Least Recently Used (LRU) 
eviction mechanism for a caching node
was found in \cite{King71}.
LRU has lower average miss rate 
compared to FIFO caching\footnote{Under FIFO caching,
the oldest item in the cache is evicted upon a cache miss.}
\cite{Belady69,Aven87,Gandolfi92}.
Numerically useful approximations  for
LRU caching nodes are found in
\cite{DS72,Fagin77,Towsley90,Jelenkovic99,Che02,Fricker12};
in particular 
the {\em expected working set miss ratio} (WS) approximation
of \cite{DS72,Fagin77} and that of 
\cite{Che02} are equivalent \cite{Jelenkovic17}.
In \cite{Jelenkovic04}, LRU caching
was studied for dependent (semi-Markov) object demand processes in a
limiting regime for certain object popularity profiles.
In \cite{Ciucu14,Cavallin15}, 
time-to-live (TTL) caching {\em networks} are studied.
Approximations for networks of 
``capacity driven" caches
are studied in \cite{Towsley10,Garetto16} (the latter adapting
the approximation of \cite{DS72,Fagin77,Che02} including
under non-LRU cache eviction policies).

Under Most Recently Used (MRU) eviction, the {\em youngest} object in
the cache is evicted upon cache miss.  
More specifically, an object is evicted under MRU when it is the subject
of a cache hit or miss (so becomes youngest) and then 
a cache miss (query for an uncached object) immediately follows.
MRU is used in cases 
where the older the object is in the cache, the more likely 
it is to be accessed
\cite{Dar96}. That is, MRU is
used when demand for hot (most popular)
objects is such that they are
not likely to be needed again soon after they are queried for,
\eg the inter-query times of hot objects are a.s. lower bounded
by a strictly positive amount,
\cf Section \ref{sec:numer}.


In this paper, we focus on single caching nodes
and present a closed-form invariant distribution 
for a standard Markov model of a generalization of LRU and MRU eviction
under the IRM.
To this end, we provide a proof  for LRU
which we will subsequently adapt.
For a Zipf popularity distribution,
numerical comparisons are made with
the simple Incremental Rank Progress (IRP)\footnote{Called CLIMB in 
\cite{Aven87}, IRP is somewhat related
to the insertion scheme based on
tandem virtual caches of ``$k$-LRU" \cite{Garetto16}.} and
and Random Eviction (RE a.k.a. random replacement, RANDOM \cite{Aven87}) methods.
Our numerical examples focus on the  range of cache-hit probabilities
for steady-state Zipf popularity distributions.
We numerically
show that the range of cache hit probabilities is smaller under MRU 
and larger under IRP compared to LRU, and conjecture that
this is true in general.
We next give 
a result for a special case of an RE caching tree-network.
The paper concludes with a summary.

\section{Background}

The generalized LRU/MRU problem we consider in the following
is similar to permutation-valued Markov chains studied
in \cite{Hendricks73,Hendricks76},
where all all objects are ranked, not just those cached.

\subsection{Markov model of
Least Recently Used (LRU) eviction policy}\label{sec:LRU}

The stationary state-space $\Rcal$ of a LRU cache is the set of
$B$-permutations of $\{1,2,...,N\}$ where $N$ is the number
of objects that could be cached and $B$ objects is the capacity of
the cache with $N>B>0$ (typically $N\gg B$) and the objects
assumed identically sized (but \cf (\ref{hit-prob-lengths})).
For $r\in\Rcal$, define $    r(k)$ as the element of $r$ in the
$\th{k}$ position.
The entries of $r$ are {\em ranked} in order  of their position in $r$:
\begin{itemize}
\item  the most recently accessed (LRU) object being
$    r(1)$, 
\item the oldest object in the cache being $    r(B)$, and 
\item uncached objects $n$ are denoted $n\not \in r$.
\end{itemize}
Note that in a transient regime, the cache may be in a state
$\not \in \Rcal$ with fewer than $B$ objects cached.

For a single node, we assume that demand process for 
object $n\in\{1,2,...,N\}$ 
is Poisson with intensity  $\lambda_n$.
The Poisson demands are assumed independent. Let 
the total demand intensity be
$\Lambda  =  \sum_{n=1}^N \lambda_n$. 
So, this is the classical ``Independent Reference Model"
(IRM) with query probabilities $p_n = \lambda_n/\Lambda$
\cite{Aven87,Gandolfi92}.

For LRU,
a cache miss of object $    r(1)$ at state $\Minv{n}(r)$ resulting in a transition
to state $r\in\Rcal$ 
occurs at rate $\lambda_{    r(1)}$, where $n\not \in r$ and  
\beqa
(\Minv{n}(r))(k) & = &  \left\{\begin{array}{cc}
n & \mbox{if $k=B$}\\
    r(k+1) & \mbox{if $k<B$}
\end{array}\right.
\eeqa
\ie $n\not\in r$ is the oldest object in the cache in state $\Minv{n}(r)$.

For LRU,
a cache hit of object $    r(1)$ at state $\Hinv{k}(r)$ resulting in 
a transition to state $r$ 
occurs at rate $\lambda_{    r(1)}$ where  $1\leq k\leq B$ and 
\beqa\label{Hinv}
(\Hinv{k}(r))(\ell) & = &  \left\{\begin{array}{cc}
    r(1) & \mbox{if $\ell =k$}\\
    r(\ell+1) & \mbox{if $\ell<k$}\\
    r(\ell) & \mbox{if $k<\ell \leq B$}
\end{array}\right.
\eeqa
\ie $r(1)$ is the $\th{k}$  youngest object in the cache in state $\Hinv{k}(r)$
and $\Hinv{1}(r)=r$.


As commonly assumed with the IRM \cite{Towsley10}, we also assume
(i) that cache misses cause
the query to be
forwarded, possibly to a server holding the requested object, and once
resolved, the object is 
reverse-path forwarded so that
caches that missed it can be updated; and (ii) the required time for
this query resolution process is negligible compared to the inter-querying
times of the caching network.

\subsection{LRU stationary invariant distribution}

The following invariant of LRU found by W.F. King in \cite{King71}.

\begin{theorem}
The unique invariant distribution of the LRU Markov chain is
\be\label{invariant-LRU}
\pi(r)  & = & 
\prod_{k=1}^B \frac{\lambda_{r(k)}}{\Lambda-\sum_{i=1}^{k-1} \lambda_{r(i)}}
\ee
for $r\in\Rcal$,
where $\forall k$, $\sum_{i=k}^{k-1} (...) \equiv 0$.
\end{theorem}

\begin{IEEEproof}
The full balance equations are: $\forall r \in \Rcal$,
\be
\lefteqn{(\Lambda-\lambda_{r(1)})\pi (r)  ~ =  } && \label{fullbal}\\
& & \sum_{n\not \in r} \lambda_{    r(1)} \pi (\Minv{n}(r)) 
+
\sum_{j=2}^B \lambda_{    r(1)} \pi (\Hinv{j}(r)).\nonumber
\ee

Under (\ref{invariant-LRU}),
for all $n\not \in     r$,
\beqa
\pi(\Minv{n}(r)) & = & 
\frac{\lambda_n} 
{\Lambda-\sum_{i=2}^{B} \lambda_{r(i)}}
\prod_{k=2}^B \frac{\lambda_{r(k)}}{\Lambda-\sum_{i=2}^{k-1} \lambda_{r(i)}}
\eeqa
Also under (\ref{invariant-LRU}),
for all $j\in\{2,3,...,B\}$,
\beqa
\pi(\Hinv{j}(r)) & =&
\prod_{k=2}^{j} \frac{\lambda_{r(k)}}{\Lambda-\sum_{i=2}^{k-1} \lambda_{r(i)}} \cdot
\frac{\lambda_{r(1)}}{\Lambda-\sum_{i=2}^{j} \lambda_{r(i)}}\\
& & ~~~ \cdot 
\prod_{k=j+1}^{B}\frac{\lambda_{r(k)}}{\Lambda-\sum_{i=1}^{k-1} \lambda_{r(i)}} 
\eeqa
Substituting into (\ref{fullbal}) and after
some term cancellation, we see that
(\ref{invariant-LRU}) satisfies (\ref{fullbal}) 
if and only if
\be
1 & = & 
 \prod_{k=3}^{B+1}\frac{\Lambda-\sum_{i=1}^{k-1} \lambda_{r(i)}}
{\Lambda-\sum_{i=2}^{k-1} \lambda_{r(i)}}  \label{fullbal-2} \\
& & ~~ + \sum_{j=2}^B
\prod_{k=3}^j
\frac{\Lambda-\sum_{i=1}^{k-1} \lambda_{r(i)}}
{\Lambda-\sum_{i=2}^{k-1} \lambda_{r(i)}} 
\cdot 
\frac{\lambda_{r(1)}}{\Lambda-\sum_{i=2}^{j} \lambda_{r(i)}} 
\nonumber
\ee
where $\prod_{k=3}^2 (...) \equiv 1$.

Regarding (\ref{fullbal-2}),    consider
the following sequence of independent random experiments
to fill the cache.
Suppose  we're given initially that the first cache entry is
$r(2)$. Now {\em sequentially},
according to the distribution 
(\ref{invariant-LRU}), object 
$r(1)$ attempts to enter the cache after $r(2)$. 
If it fails to enter in the $\th{k}$ attempt, then $r(k+2)$ is placed
in the cache instead and $r(1)$ tries again.
The summand of (\ref{fullbal-2}) 
with $j=2$
is the probability that $r(1)$ enters in the second position right
after $r(2)$:
$\lambda_{r(1)}/(\Lambda-\lambda_{r(2)})$.
Generally, the summand for $j\in\{2,3,...,B\}$ is the probability 
$r(1)$ enters in the $\th{j}$ position (after having failed to
enter in one of the more highly ranked ones).
The first term of the right-hand-side of
(\ref{fullbal-2}) is the probability $r(1)$ fails to enter the cache.
So, (\ref{fullbal-2}) must generally hold by the law of total probability.

Finally, since the stationary
LRU Markov chain is irreducible on $\Rcal$, 
there is a unique  invariant.
\end{IEEEproof}

This result was  generalized in 
\cite{ST01} to add object-dependent  
insertion probabilities interpreted as access costs.
Also note that, generally,
the LRU Markov chain is neither time-reversible nor quasi-reversible
\cite{Serfozo98}.
Obviously, 
more popular objects (larger $\lambda$) are more likely stored, and
the LRU invariant is uniform in the special case
that all the mean querying rates $\lambda_n$ are the same.
Finally, by PASTA,
the stationary hit probability of object $n$ in a LRU cache is 
\beqa
h_n & = & \sum_{r~:~n\in r} \pi(r),
\eeqa
where the approximations of hit probabilities in \cite{Che02,Fricker12} 
are obviously substantially simpler to compute.

\subsection{Incremental Rank Progress (IRP or ``CLIMB" \cite{Aven87}) upon query}\label{sec:IRP}

Under LRU, a query for any object $n$ results in it being ranked first
in the cache.
One can also consider slowing the ``progress through the ranks" of objects
as they are queried, leading to some obvious trade-offs with LRU: Slowing 
progress would mean
less popular content does not enter the cache at first rank, but also
more popular content will take longer to reach the cache. 
Such issues are important when there are dynamic
changes/churn in objects cached and their popularity.  

Under an Incremental Rank Progress (IRP) caching mechanism,
a query for object $n$ results in its  rank improved by just one
(or zero if the object
is already ranked first), \ie for $1\leq k\leq B-1$, $r\in \Rcal$,
\beqa
(T_k(r))(\ell) & = &  \left\{\begin{array}{cc}
    r(k) & \mbox{if $\ell =k+1$}\\
    r(k+1) & \mbox{if $\ell =k$}\\
    r(\ell) & \mbox{else}
\end{array}\right.
\eeqa
where the transition $T_{k}(r) \rightarrow r$ with rate $\lambda_{r(k)}$.
Missed objects  enter the cache at lowest rank, \ie
for $n\not \in r$, define
\beqa
(S_{n}(r))(\ell) & = &  \left\{\begin{array}{cc}
    r(k) & \mbox{if $\ell <B$}\\
    n & \mbox{if $\ell =B$}
\end{array}\right.
\eeqa
where the transition $S_{n}(r) \rightarrow r$ 
occurs with rate $\lambda_{r(B)}$.
The invariant for IRP is found in \cite{Aven87} and
can be immediately shown using detailed balance.

\begin{theorem}\label{IRP-single-node}
IRP is time-reversible with 
unique stationary invariant  
\be\label{invariant-IRP}
\pi(r) &  =  & \frac{\prod_{k=1}^B \lambda_{r(k)}^{B+1-k}}
{\sum_{r'\in\Rcal}\prod_{k=1}^B \lambda_{r'(k)}^{B+1-k}}.
\ee
\end{theorem}


\subsection{Random Eviction  (RE or ``RANDOM" \cite{Aven87}) 
upon cache miss without cache rankings}\label{sec:RE}

Suppose that 
a cache miss of object $n$ at state $\Minv{\ell,n}(r)$ results in a transition
to state $r\in\Rcal$   at rate $B^{-1}\lambda_{n}$, where
$n\in r$, $n\not\in\Minv{\ell,n}(r)$, 
$\ell \in\Minv{\ell,n}(r)$, and $\ell \not\in r$.
That is, a cache miss for object $n$ results in $n$ inserted into the cache
and evicting of an object $\ell$  selected 
uniformly at random from the cache.
The cache state $r$ does not change if a cache hit occurs.
The stationary state-space
$\Rcal$ is the set of $B$-{\em combinations} of $N$ different objects.
The following invariant for RE is
also found in \cite{Aven87}
and can also be immediately shown by detailed balance.

\begin{theorem}
The RE Markov chain is time-reversible with 
unique stationary invariant distribution  
\be\label{invariant-RE-NPTR}
\pi(r)  & = & \frac{\prod_{n\in r} \lambda_{n}}
{\sum_{r'\in\Rcal} \prod_{n\in r'} \lambda_{n}}.
\ee
\end{theorem}

\subsection{Aggregate cache-hit rates}

Define the aggregate hit rate for a
caching discipline as
\be
H & := & \sum_{n=1}^N h_n p_n  = \sum_{n=1}^N h_n \frac{\lambda_n}{\Lambda},
\label{aggr-hit-rate}
\ee
\ie the probability that a query is a cache hit.
This is a single criterion that can be used to compare
different caching disciplines. Typically $H$ is largest
for LRU  eviction under the IRM.
Note that under the IRM, 
by PASTA and Fubini's theorem 
the following holds for all of the above 
capacity-driven caching disciplines,
\be
\sum_{n=1}^N h_n  =  \sum_{n=1}^N \sum_{r\in \Rcal:n\in r} \pi(r) 
 =   \sum_{r\in \Rcal} \pi(r) B 
 =   B. \label{sum-hit-rates}
\ee

\subsection{Considering objects with different lengths}

To account for objects of different lengths for capacity-driven
caches (with ranked objects) like LRU, simply
consider a ``complete-rankings"  LRU variation,
where the ranking of {\em all} objects 
is maintained whether the objects are cached
or not. That is, the state-space $\Rcal$ is now 
the set of permutations of {\em all} $N$ objects.

\begin{corollary}
The unique stationary invariant 
$\pi$ of complete-rankings LRU is
(\ref{invariant-LRU}) with $B$ replaced by $N$.
\end{corollary}

Additionally consider the different sizes $\ell_n$ bytes of objects $n$,
where the cache capacity $B$ is in bytes.
The number of objects in the cache is given by
\beqa
K(r) & = & \max \{ K ~|~ \sum_{k=1}^K \ell_{r(k)} \leq B,~ 1\leq K\leq N\}.
\eeqa
So, the hit probability of object $n$ when the objects are of variable length is
\be\label{hit-prob-lengths}
h_n & = & \sum_{r~:~r(n)\leq K(r)} \pi(r).
\ee
See the byte-hit  performance metric of \cite{Krunz04}.

\section{Most Recently Used (MRU) eviction}\label{sec:MRU}

Again define the state-space $\Rcal$ as the
set of $B$-permutations of $\{1,2,...,N\}$.
Under MRU 
\cite{Dar96,Krunz04},
a cache hit of object $    r(1)$ at state $\Hinv{k}(r)$ resulting in 
a transition to state $r$ 
occurs at rate $\lambda_{    r(1)}$ where  $1\leq k\leq B$ and 
$(\Hinv{k}(r))(\ell) $ is given by (\ref{Hinv}) as LRU.
But for MRU, a cache miss of object $    r(1)$ at state $\Minv{n}(r)$ resulting in a transition
to state $r\in\Rcal$ 
occurs at rate $\lambda_{    r(1)}$, where $n\not \in r$ and  
\beqa
(\Minv{n}(r))(k) & = &  \left\{\begin{array}{cc}
n & \mbox{if $k=1$}\\
    r(k) & \mbox{if $k>1$}
\end{array}\right.
\eeqa
\ie $n\not\in r$ is the 
{\em youngest} object in the cache in state $\Minv{n}(r)$.

\begin{theorem}
The unique invariant distribution of the MRU Markov chain
is, for  $r\in\Rcal$,
\small
\be
\pi(r) ~= \hspace{2.5in} & & \nonumber\\
\frac{\lambda_{r(1)}}{\Lambda}\cdot\frac{1}{\binom{N-1}{B-1}} 
\prod_{k=2}^{B-1} \frac{\lambda_{r(k)}}
{\Lambda-\sum_{i=1}^{k-1} \lambda_{r(i)} - \sum_{n\not\in r}\lambda_n}.
& &
\label{invariant-MRU} 
\ee
\normalsize
\end{theorem}

\begin{IEEEproof}
The full balance equations are as for LRU but with a different
definition for $\Minv{n}$.

Let $\Lambda_{r} = \Lambda-\sum_{n\not\in r}\lambda_n$.
By substituting (\ref{invariant-MRU}) into the full
balance equations (and moving the cache-miss terms to the left-hand side), 
we get that 
(\ref{invariant-MRU}) satisfies the full balance equations if and only if
\small
\be
1 & = &  
\frac{1}{\Lambda_{r}-\lambda_{r(1)}}\left(\lambda_{r(1)} 
\sum_{j=2}^{B-1} 
\prod_{k=2}^{j-1} \frac{\Lambda_{r} - \sum_{i=1}^{k-1} \lambda_{r(i)}}
{\Lambda_{r} - \sum_{i=2}^{k} \lambda_{r(i)}}  \right. \nonumber\\
&   &   \left. ~~+
\lambda_{r(B)} 
\prod_{k=2}^{B-1} \frac{\Lambda_{r} - \sum_{i=1}^{k-1} \lambda_{r(i)}}
{\Lambda_{r} - \sum_{i=2}^{k} \lambda_{r(i)}} \right) \nonumber\\
& = & \sum_{j=2}^{B-1} 
\left(\prod_{k=2}^{j-1} \frac{\Lambda_{r} - \sum_{i=1}^{k} \lambda_{r(i)}}
{\Lambda_{r} - \sum_{i=2}^{k} \lambda_{r(i)}} \right)
\frac{\lambda_{r(1)}}
{\Lambda_{r} - \sum_{i=2}^{j} \lambda_{r(i)}}  \nonumber\\
&   &    ~+
\prod_{k=2}^{B-1} \frac{\Lambda_{r} - \sum_{i=1}^{k} \lambda_{r(i)}}
{\Lambda_{r} - \sum_{i=2}^{k} \lambda_{r(i)}} 
\label{fullbal-MRU} 
\ee
\normalsize
where $\prod_{k=2}^1 (...) \equiv 1$.

Regarding (\ref{fullbal-MRU}),    consider
the following sequence of independent random experiments
to determine the position of object $\lambda_{r(1)}$ 
when filling the cache, {\em given}
that only objects  $\in r$ will be chosen and that
$\lambda_{r(2)}$ has already been chosen first.
$\lambda_{r(1)}$ is chosen on the first try with
probability $\lambda_{r(1)}/(\Lambda_{r}-\lambda_{r(2)})$,
otherwise $\lambda_{r(3)}$ enters the cache - this is the
summand of (\ref{fullbal-MRU}) with $j=2$. 
Generally, the $\th{j}$ summand is the probability that $\lambda_{r(1)}$
enters the cache on the $\th{(j-1)}$ try, otherwise object
$\lambda_{r(j+1)}$ is placed in the cache.
The final term  of
(\ref{fullbal-MRU}) is the probability $r(1)$ fails to enter the cache
before the last ($\th{B}$) position,
because in the penultimate choice only objects $r(B)$ and $r(1)$ remain, \ie
$\lambda_{r(B)}=\Lambda_{r} - \sum_{i=1}^{B-1}\lambda_{r(i)}$.
So, (\ref{fullbal-2}) must generally hold by the law of total probability.

Finally, since the stationary
LRU Markov chain is irreducible on $\Rcal$, 
there is a unique  invariant.
\end{IEEEproof}

Note that it's easily directly verified that (\ref{invariant-MRU}) satisfies
(\ref{fullbal}) for the cases  $B=2$ and $B=3$, \eg for $B=3$ and $N=4$,
$$\pi(r) = \lambda_{r(1)}\lambda_{r(2)}/(3\Lambda(\lambda_{r(2)}+\lambda_{r(3)})).$$
To interpret (\ref{invariant-MRU}): $\lambda_{r(1)}$ is chosen
with probability $\lambda_{r(1)}/\Lambda$; then 
the remaining $B-1$ objects in $r$ are chosen  from
the remaining $N-1$ objects
uniformly at random with probability 
$\binom{N-1}{B-1}^{-1}$; finally, the order of the
remaining items $\lambda_{r(2)}, \lambda_{r(3)}, ...$ are
determined as the LRU invariant distribution
(\ref{invariant-LRU}).


Finally, we make an observation about cache-hit probabilities under MRU
eviction.
Consider a MRU cache under the IRM 
that is ``synchronized" so that a query for object $n$ occurs at time 0.
Thus, immediately thereafter, $n$ is the MRU object in the cache.
The next query for object $n$ will be at time $T_n\sim\exp(\lambda_n)$.
Again, under MRU eviction, the only way an object $n$ is evicted is when a cache
miss occurs immediately after a query for $n$, \ie a cache miss when
$n$ is the MRU object.
So, the stationary hit probability $h_n$ of object $n$  equals
the probability that a hit occurs at time $T_n$, which is 
\begin{itemize}
\item the probability that no other queries occurred in the interval $(0,T_n)$ 
plus
\item  the probability that a query does occur in $(0,T_n)$ and the first
such query is a hit.
\end{itemize}
Thus, we can write $\forall n$,  
\small
\beqa
h_n~ = \hspace{2.5in} & & \\
 \E\left( 
\mbox{e}^{-T_n \sum_{j\not = n} \lambda_j}
+ (1- \mbox{e}^{-T_n \sum_{j\not = n} \lambda_j}) \sum_{j\not = n}
\frac{\lambda_j h_{j|n}}{\sum_{i\not =  n} \lambda_i} 
\right) , 
&  &
\eeqa
\normalsize
where $h_{j|n}$ is the probability that a query is a hit on $j$ 
given that object $n$ is MRU.
We have therefore shown the following.

\begin{proposition}
For a MRU-eviction cache under the stationary IRM: $\forall n$, 
$h_n  =  p_n + \sum_{j\not = n} p_j h_{j|n} ~=~ \sum_j p_j h_{j|n}$,
where $p_j = \lambda_j / \sum_i \lambda_i$ and $h_{j|j}=1$;
equivalently, a kind of balance equation: $\forall n$,
\beqa
\sum_j p_j h_{n|j} & = & \sum_j p_j h_{j|n}.
\eeqa
\end{proposition}

\section{Generalization of LRU and MRU}\label{sec:variations}

``$k^{\rm th}$ Recently Used" ($k$RU) is a simple
generalization of LRU and MRU wherein
object $r(k)$, for some fixed $k\in\{1,2,...,B\}$,
is evicted upon cache miss; otherwise cache insertion (at rank $1$) upon
misses and promotion (to rank $1$) and demotions (by $1$) upon hits 
are the same as both MRU and LRU. That is, $B$RU is LRU and
$1$RU is MRU.

\begin{corollary}\label{cor:kRU}
The invariant distribution of $k$RU is
\small
\be
\pi(r)  & = &
\prod_{j=1}^{k} \frac{\lambda_{r(j)}}{\Lambda - \sum_{i=2}^j\lambda_{r(i)}}
\label{invariant-kRU}\\
& & \times \frac{1}{\binom{N-k}{B-k}}
\prod_{j=k+1}^{B-1} \frac{\lambda_{r(j)}}
{\Lambda-\sum_{i=1}^{j-1} \lambda_{r(i)} - \sum_{n\not\in r}\lambda_n}.
\nonumber
\ee
\normalsize
\end{corollary}

\section{Numerical results for small $N,B$}\label{sec:numer}

\begin{figure}[h]
\begin{center}
\includegraphics[width=\columnwidth]{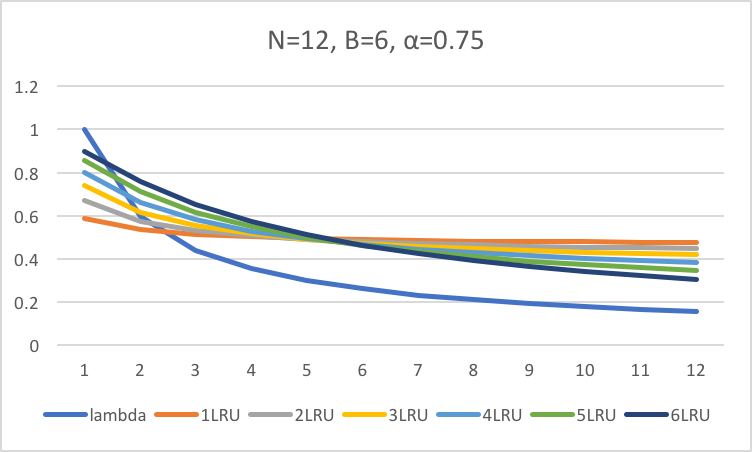}
\caption{$k$RU cache hit probabilities $h_n$ and popularity $\lambda_n$
versus object index $n$ 
for a cache of
size $B=6$, $N=12$ objects, and Zipf popularity parameter
$\alpha=0.75$, where LRU$=6$RU and MRU=$1$RU }\label{fig:kRU_N12B6A075}
\end{center}
\end{figure}

In this numerical study, we directly computed 
the invariants $\pi$ by generating all possible object permutations
representing cache state by the Steinhaus-Johnson-Trotter algorithm.
So, we considered only small values for the number of objects
and the cache size. 
Figure  \ref{fig:kRU_N12B6A075} 
is representative of our numerical study on cache-hit probabilities 
using a Zipf popularity model
$\lambda_n = n^{-\alpha}$ for with $\alpha=0.75$
(see Table 1 of  \cite{Shenker99}) and most popular object
indexed 1 with normalized rate $\lambda_1=1$.

$k$RU with $1<k<B$ gives hit-probability performance between
MRU ($k=1$) and LRU  ($k=B$).
That is, one can see that the range of hit probabilities
for LRU is larger than that of MRU. 

\begin{figure}[h]
\begin{center}
\includegraphics[width=\columnwidth]{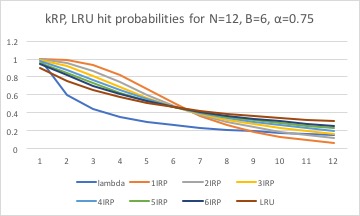}
\caption{$k$RP (with cache entry upon cache miss)  and LRU
cache hit probabilities $h_n$ and popularity $\lambda_n$
versus object index $n$
for a cache of
size $B=6$, $N=12$ objects, and Zipf popularity parameter
$\alpha=0.75$}\label{fig:kRP_N12B6A075}
\end{center}
\end{figure}

Figure \ref{fig:kRP_N12B6A075} shows the results of a
typical simulation study
of $k$RP with cache entry at lowest rank $B$ upon cache miss
compared to LRU. Note that $k$RP has greater range of hit probability
values than LRU.
We postulate that generally for Zipf popularity
distributions, the range of hit probabilities of IRP is larger
than those of LRU which is larger than those of MRU.

\begin{figure}[h]
\begin{center}
\includegraphics[width=\columnwidth]{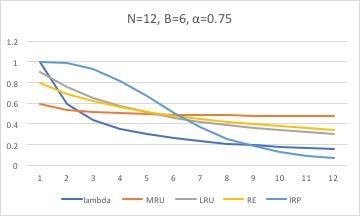}
\caption{Cache hit probabilities $h$ versus popularity $\lambda$
for a cache of
size $B=3$, $N=12$ objects, and Zipf popularity parameter
$\alpha=0.75$}\label{fig:N12B3A075}
\end{center}
\end{figure}

For the example of Figure \ref{fig:N12B3A075},
RE has a range of hit probabilities between MRU and LRU.
Recall 
(\ref{sum-hit-rates}), \ie
 that the sum of the 
stationary hit probabilities is the same for all of these caching disciplines
under the IRM

Though our derivations herein are
for the IRM,
MRU may out-perform LRU for non-Poisson arrivals
in terms of aggregate hit rate (\ref{aggr-hit-rate}).
Recall mention in the Section 1 that MRU is
used  when demand for hot (most popular)
objects is such that they are
not likely to be needed again soon after they are queried for.
Consider the case where inter-query times are lower bounded
by a constant $D$. Specifically, inter-query times
equal $D$ plus an exponentially
distributed quantity, such that $D=0$ corresponds to 
the IRM
(here with intensities following
a Zipf popularity distribution).
In Table \ref{table:MRU-dominates}, we see
that LRU has best aggregate hit rate under IRM
(mean hit rate increases with $k$ when
$D=0$), while   MRU is best when $D=1,2$
(mean hit rate decreases with $k$ when
$D=2$).

\begin{table}
\centering
\begin{tabular}{l|ccc}
$k$     & $D=0$ & $D=1$ & $D=2$ \\ \hline
1 (MRU)  & 0.52  & 0.4578  & 0.45 \\
2       & 0.54  & 0.4213  & 0.40 \\
3       & 0.56  & 0.4014  & 0.35 \\
4       & 0.58  & 0.4026  & 0.31 \\
5       & 0.60  & 0.4187  & 0.29 \\
6 (LRU)  & 0.62  & 0.4423  & 0.29
\end{tabular}
\caption{$k$RU aggregate hit rate (\ref{aggr-hit-rate}) for
$N=12$ objects, cache of capacity $B=6$ objects,
and Zipf popularity distribution with exponent  $\alpha=0.75$  
($D=0$ corresponds to the IRM).} \label{table:MRU-dominates}
\end{table} 

\section{Discussion: Networks of RE caches}

\begin{figure}[h]
\begin{center}
\includegraphics[width=\columnwidth]{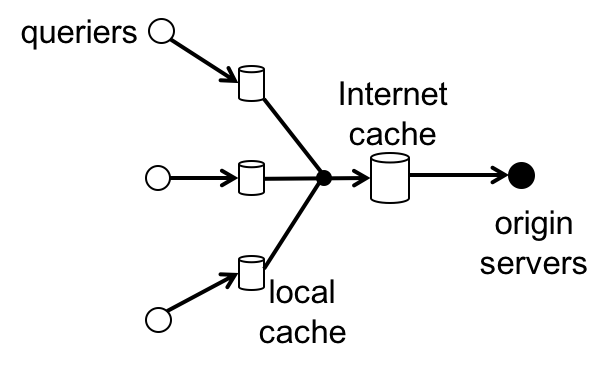}
\caption{A tree-network of caching nodes that feeds forward cache misses
with assumed independent local caches}\label{fig:caching-net}
\end{center}
\end{figure}

The performance of Markovian networks of such capacity-driven caches
are approximated 
in \eg \cite{Towsley10,Garetto16}.
To illustrate the difficulties with capacity-driven
caching networks,
now consider the simplest ones based on RE.
Though  RE caches are time-reversible,
a tree of independent local caches whose collective
query-misses are forwarded to an Internet cache (also
running RE, see Figure  \ref{fig:caching-net}),
is not time-reversible and its non-local nodes do
not operate under the IRM.  To see why it's not
time-reversible,
consider a cache miss of object $n$ of local cache $q$ 
of size $B_q$ in state $r_q$, so that object $n_q$ is evicted,
and suppose it's also a miss on the Internet cache of size $b$ in state $R$,
so that object $n$ is evicted; this can be reversed with one query
(so that states $r_q$ and $R$ are restored) only if $n=n_q$.

The following result is for the very special case
that the Internet cache holds only one object. 

\begin{proposition}\label{prop:RE-network}
The invariant distribution $\pi$ of  the 
network Figure \ref{fig:caching-net} with RE caching 
and $b=1$ satisfies 
\be \label{invariant}
\pi (R | \ur) & = &  
 \frac{\sum_{q} {\bf 1}\{R\in r_q\} \Lambda_{q,\overline{r}_q}/B_q}
 {\sum_{q} \Lambda_{q,\overline{r}_q}}
\ee
where 
\beqa
\Lambda_{q,\overline{x}} & = & 
\sum_{\ell\not\in x}\lambda_{q,\ell},
\eeqa
$\sum_{\emptyset} (...)\equiv 0$, and  indicator
${\bf 1}X=1$ if $X$ is true otherwise $=0$.
\end{proposition}

\begin{IEEEproof}
For $n\in r_q,m\not \in r_q$, let $\delta_{q-n+m}\ur$ be
$\ur$ but with $n$ in $r_q$ replaced by $m$.
Similarly define $\delta_{-n+\ell} R$.
The full balance equations are
\small
\beqa
\lefteqn{\pi(\ur,R) \sum_{q,m:m\not \in r_q} \lambda_{q,m}} & & \\
& = &
\sum_{q,m,n:m\not\in r_q;n\in r_q\cap R} \pi(\delta_{q-n+m}\ur,R)\frac{\lambda_{q,n}}{B_q}   +
\\
& & \sum_{q,m,n,\ell:m\not\in r_q;n\in r_q\cap R;\ell \not\in R} 
\pi(\delta_{q-n+m}\ur,\delta_{-n+\ell} R)\frac{\lambda_{q,n}}{B_q b}  
\eeqa
\normalsize
Dividing by $\pi(\ur,R)=\pi(R|\ur)\prod_q\pi(r_q)$ and then substituting the 
stationary joint distribution of the independent RE local caches
 (\ref{invariant-RE-NPTR})
into the full balance equations gives:
$\forall \ur,R,$
\small
\beqa
\pi(R|\ur)\sum_{q,m:m\not \in r_q} \lambda_{q,m}  
~ = ~
\sum_{q,m:m\not\in r_q} 
\frac{\lambda_{q,m}}{B_q}   \times \hspace{1.5in}\\
\sum_{n\in r_q\cap R}\left(\pi(R|\delta_{q -n+m}\ur)
+ \frac{1}{b}\sum_{\ell \not \in R}
\pi(\delta_{+l-n}R|\delta_{q-n+m}\ur)\right)    \hspace{0.70in}
\eeqa
\normalsize
For the special case of $b=1$, \ie $R$ ($=n$) is a single object, 
we get that the 
right-hand-side simplifies to
\small
\beqa
\lefteqn{\sum_{q,m:m\not\in r_q} 
\frac{\lambda_{q,m}}{B_q} {\bf 1}\{R\in r_q\}}&&\\
\lefteqn{\times \left(\pi(R|\delta_{q-R+m}\ur)
+\sum_{\ell \not = R}
\pi(\ell|\delta_{q-R+m}\ur)\right)} & &  \\
&=& \sum_{q,m:m\not\in r_q} 
\frac{\lambda_{q,m}}{B_q} {\bf 1}\{R\in r_q\} 
 ~=~   \sum_q \frac{\Lambda_{q,\overline{r}_q}}{B_q} {\bf 1}\{R\in r_q\} 
\eeqa
\normalsize
The invariant is unique since $(R,\ur)$ is irreducible.
\end{IEEEproof}
~\\


In steady state, $R\subset \cup_q r_q$ a.s., \ie if
$\forall q, ~R \not \in  r_q$ then $\pi(R|\ur)=0$.
Note that (\ref{invariant}) is the eviction probability of object $R$
upon local cache miss in local cache state $\ur$.
An individual
RE cache $r$ is not quasi-reversible 
since the miss rates (``departures"),
$\frac{1}{\pi(r)}\sum_{m\not \in r,n\in r} \lambda_n \pi (\delta_{-n+m}(r))$
depend on the state $r$. Though quasi-reversibility is not a
necessary condition \cite{Serfozo98}, Proposition \ref{prop:RE-network}
shows that RE networks generally do  not have product-form invariants.
More specifically, one can identify 
the incident mean rate of queries for object $n$ to the Internet cache,
$\widehat{\lambda}_n := \sum_q  \lambda_{q,n}(1-h_{q,n})
= \sum_q  \lambda_{q,n}\sum_{r_q:n\not\in r_q} \pi(r_q)$,
where $1-h_{q,n}$ is the stationary miss probability of local cache $q$
for object $n$ under RE\footnote{In this way, one can easily  identify
the ``flow-balance equations" 
for more general caching networks \cite{Towsley10}.}.
According to this proposition, $\pi(R)$ does not depend on 
the $\widehat{\lambda}_n$
in the way the IRM invariant $\pi(r_q)$ depends on the  $\lambda_{q,n}$
 in (\ref{invariant-RE-NPTR}),
\ie $\pi(R)=\sum_{\ur} \pi(R|\ur)\pi(\ur)=\sum_{\ur}\pi(R|\ur)\prod_q
\pi(r_q) \not = \widehat{\lambda}_R/\sum_n \widehat{\lambda}_n$.
Finally note that, since the capacity of 
the Internet cache is one object ($b=1$), 
it could obviously be operating any eviction policy.


\section{Summary}\label{sec:summary}

In this paper, under the IRM, a closed-form
expression for the invariant distribution was derived 
for a caching node using $k$RU eviction.
Numerically, it was shown that under IRM and Zipf popularity
distributions for the data objects, the range of cache-hit 
probabilities of the data objects under
IRP caching is larger than LRU,
which is larger than RE, which is larger than MRU
(also, a non-IRM example was given where MRU had higher
aggregate hit rate than LRU).
Finally, the invariant distribution of a special case
of a Markovian RE caching tree-network was also derived.

\bibliographystyle{IEEEtran}
\bibliography{../latex/cache,../latex/quasi-rev,../latex/sim_ann,../latex/wiley}

\end{document}